%% file: Third_Submission_2013_05_12_-_performance_limits_of_MISO_scalar_coding.tex
\documentclass[12pt,journal,onecolumn,romanappendices]{IEEEtran}
\usepackage{calc}
\usepackage{amsmath,epsfig,psfrag,graphicx}
\usepackage{amsfonts}
\usepackage{bbding}
\usepackage{flushend,cuted}
\usepackage{setspace}
\usepackage{hyperref}

\input{rv_defs}

\newtheorem{lemma}{Lemma}

\newtheorem{proposition}{Proposition}

\newcommand{\bh}{{\mathbf{h}}}

\newcommand{\eps}{\varepsilon}

\newcommand{\SNR}{\mathrm{SNR}}

\begin{document}

\title{On the Performance Limits of Scalar Coding Over MISO Channels}


\author{Elad~Domanovitz and Uri~Erez,~\IEEEmembership{Member,~IEEE}
\thanks{This work was supported in part by the Israel Science Foundation, grant ISF1234/09 and  by the Binational Science Foundation, grant
BSF2008455. The results of this work appears in part in \url{http://www.eng.tau.ac.il/\~uri/elad_domanovitz_Msc.pdf}.}
\thanks{E.~Domanovitz and U.~Erez are with the Department of
Electrical Engineering - Systems, Tel Aviv University, Ramat Aviv,
69978, Israel (Email: \{domanovi,uri\}@eng.tau.ac.il).}
}

\maketitle

\renewcommand{\thefootnote}{}
\renewcommand{\thefootnote}{\arabic{footnote}}

\begin{abstract}
The performance limits of scalar coding for multiple-input single-output channels are revisited in this work.
By employing randomized beamforming, Narula et al. demonstrated that the loss of scalar coding is universally bounded by $\sim2.51$ dB (or 0.833 bits/symbol) for any number of antennas and channel gains. In this work, by using randomized beamforming in conjunction with space-time codes, it is shown that the bound can be tightened to $\sim1.1$ dB (or 0.39 bits/symbol).
\end{abstract}

\section{Introduction}
We consider a multiple-input single-output (MISO) system, where a transmitter equipped with $M$ antennas
communicates with a receiver equipped with a single antenna. We further consider an open-loop mode of operation, i.e., the transmitter is assumed to have no knowledge of the channel. The receiver is assumed to have perfect channel state information. The complex baseband received signal at time $t$ is
\begin{eqnarray}
y^{\rm{rec}}_{t} & = & \boldsymbol{h}^T\boldsymbol{x}_t+{n}^{\rm{rec}}_{t}
\label{eq:basicScalarModel}
\end{eqnarray}
where $\boldsymbol{x}_t=\left[{x}_{1,t},....,{x}_{M,t}\right]^T$ is the input vector and ${n}^{\rm{rec}}_{t}$ is i.i.d. circularly-symmetric complex white Gaussian noise with power $N_0$. The components of $\boldsymbol{{x}}_t$ are assumed to be uncorrelated between antennas, each with power
\begin{eqnarray}
E\left[|{x}_{i,t}|^2\right]=\frac{\eps_s}{M}, \nonumber
\end{eqnarray}
where $\eps_s$ is the total transmit power.
Under these assumptions, the mutual information is maximized when $\boldsymbol{{x}}$ is i.i.d. circularly-symmetric Gaussian, yielding (see, e.g., \cite{telatar})
\begin{eqnarray}
I_{\rm OPT}(\SNR) 
=\log_2\left(1+\frac{\SNR||\boldsymbol{h}||^2}{M}\right),
\end{eqnarray}
where
\begin{eqnarray}
\SNR=\frac{\eps_s}{N_0} \nonumber.
\end{eqnarray}
In the sequel we refer to $I_{\rm OPT}$ as the WI mutual information. Without loss of generality we assume that $||\boldsymbol{h}||^2/M=1$ and hence
\begin{eqnarray}
I_{\rm OPT}(\SNR)= \log_2\left(1+\SNR\right).
\label{eq:OPTwithChanNorm1}
\end{eqnarray}

We emphasize that $I_{\rm OPT}(\SNR)$ depends on the channel vector only through its norm $||\boldsymbol{h}||$. Thus, $I_{\rm OPT}$ is the maximal possible rate with isotropic transmission. We will refer to a modulation scheme that maintains this property as a ``norm dependant only'' (NDO) modulation scheme.

In the sequel, we discuss linear modulation schemes that convert a MISO channel to a (possibly time-varying) single-input single-output (SISO), over which a scalar code is utilized. Following \cite{Narula}, we refer to such a coding and modulation approach as ``scalar coding".
A measure for the performance of a scalar coding scheme will be its mutual information.

We assume that the MISO channel remains constant throughout transmission of a codeword and no statistical assumptions enter the analysis. Rather, we study the ``worst-case" (WC) mutual information loss of scalar coding schemes. That is, for a given number of transmit antennas and a given scalar coding scheme, we consider the {\em maximum gap} in mutual information, over all channels $\bh$ for which $||\boldsymbol{h}||^2/M=1$ as well as all values of $\SNR$, between $I_{\rm OPT}$ and that achieved by the considered scheme.

The problem formulation is of interest in various communication setting. For instance, consider a transmitter equipped with an antenna array, sending a common message (i.e., multicasting) to a number of users, where each user is equipped with a single antenna and where all channels coefficients are constant. In such a setting (for sufficiently large channel coherence time), not much overhead would be required to allow the transmitter to obtain knowledge of the channels via feedback links and thus one could work in a closed-loop mode.
For a single receiver this would allow to use beamforming. However, when the number of receivers is large (and their channel vectors are uncorrelated), it is not hard to see that channel state information buys us little, and isotropic transmission is optimal in the limit of many users. Thus, multicast (broadcast of a common message) in such a setting essentially reduces to open-loop transmission.

The NDO property allows to obtain a bound on the performance loss (measured in dB or bits/channel use) w.r.t. to the optimal possible performance which is independent of channel statistics. For instance, in an open-loop single user scenario the outage capacity of scalar coding will be lower bounded by the optimal outage capacity minus the obtainable universal bound irrespectively of the outage probability as well as the channel statistics.

In \cite{Narula} it was shown that the loss of mutual information incurred by scalar coding is no greater than $\sim2.51$ dB. This was established by using randomized beamforming to transform the MISO channel into a scalar one. Since the publication of \cite{Narula}, great progress has been made in approaching the limits of MISO channels using scalar coding. Most notably, the gap was shown by Alamouti \cite{Alamouti} to be zero for the case of two transmit antennas. For more than two antennas, many extensions of Alamouti
modulation have been developed, but none allow to achieve the information-theoretic limit of (\ref{eq:OPTwithChanNorm1}).
In this work, we study to what extend the space-time coding methods developed since the publication of \cite{Alamouti} allow to tighten the bounds obtained by
Narula et al. The tightest bound we obtain, which holds for any number of antennas, and which we believe is the tightest bound available to date, is obtained using Alamouti modulation in conjunction with an extension of randomized beamforming to two dimensions. The scheme is a slight extension of the modulation scheme proposed in \cite{TROMBI} (where it was named TROMBI). Specifically, we show that the gap-to WI mutual information incurred by scalar coding is no greater than $\sim1.1$ dB.

The paper is organized as follows. In Section~\ref{sec:scalar coding}, the framework considered for transforming MISO channels to SISO channels (i.e., scalar coding) is described. In Section~\ref{sec:Review of Known Results} we review known scalar coding techniques. In Section~\ref{sec:RBF} we describe a generalized TROMBI modulation scheme which projects the MISO to two ``virtual'' channels. We analyze its performance, and derive an upper bound on the associated gap to WI mutual information.
The paper concludes with Section~\ref{sec:conc}.

\section{scalar coding}
\label{sec:scalar coding}
Scalar-coded antenna systems might be viewed as antenna systems where a scalar code is used in conjunction with linear processing; see \cite{Narula} for a detailed account. In essence, by linear pre/post processing, the MISO channel is transformed into a SISO one, as depicted in Fig.~\ref{fig:scalarScheme}, and a scalar code is used for transmission over the resulting SISO channel.
Such an approach is attractive due to its modularity, i.e., coding and modulation are effectively decoupled and standard coding techniques developed for SISO channels may readily be used.

Perhaps the most celebrated example of a linear modulation scheme, that transforms a MISO channel into a scalar one, is that of Alamouti \cite{Alamouti} modulation.\footnote{Henceforth, we refer to the Alamouti scheme as a modulation scheme rather than a coding scheme as is more
common in the literature. We do so not only because it better reflects the nature of the scheme but also to differentiate it from the scalar code that is applied to the resulting scalar channel. We adopt this nomenclature when referring to all space-time ``codes" in the sequel.}
For the case of two transmit antennas, Alamouti modulation transforms the MISO channel
via linear pre/post processing into two orthogonal channels (for transmitting two data symbols over two channel uses), resulting in an AWGN scalar channel with mutual information equal to $I_{\rm OPT}$.

Numerous extensions of Alamouti modulation have been developed for systems equipped with more than two transmit antennas. Well known examples for quasi-orthogonal space-time block codes (QOSTBC) which will be discussed in the sequel are ABBA \cite{Tirkkonen}, extended Alamouti (EA) \cite{Jafarkhani} and the Papadias-Foschini (PF) \cite{Papadias&Foschini}). However, all of these extensions suffer some loss in mutual information and do not achieve $I_{\rm OPT}$. Another extension of Alamouti modulation which sums two identical copies of the Alamouti modulation together with pseudo-random phase was introduced in \cite{TROMBI} and was referred to as TROMBI (or STTD-PHOP in subsequent works) modulation and will also be discussed in the sequel.

In~\cite{Tarokh}, Alamouti modulation was generalized to the family of orthogonal space-time block codes (OSTBC), a family of modulation schemes that retains the orthogonality property but at the price of a loss in the number of utilized degrees of freedom (i.e., reduced symbol rate). Specifically, it has been shown in \cite{Liang} that the maximal possible symbol rate that a complex orthogonal design for $M$ transmit antennas can achieve is $\frac{m+1}{2m}$ where $m$ is a natural number such that $M=2m$ or $M=2m-1$. As this means that the maximal symbol rate is strictly smaller than one, it follows that the loss in mutual information is unbounded at high SNR. At asymptotically low SNR, on the other hand, it is readily seen that OSTBC schemes approach optimal performance. That is, the ratio of the mutual information achievable with OSTBC to $I_{\rm OPT}$ goes to one, as the SNR goes to zero. We further note that OSTBC modulation schemes are NDO.

Prior to the development of orthogonal and quasi-orthogonal modulation schemes, Narula et al. extensively studied scalar coding schemes for MISO channels in \cite{Narula}. In particular, they studied the information-theoretic limits of systems (having any number of transmit antennas) employing isotropic randomized beamforming (IR-BF). It is observed in \cite{Narula} that by applying IR-BF, the MISO channel is converted to a SISO channel (with time-varying SNR) with NDO mutual information.
Of particular relevance to the present paper is the result of \cite{Narula}, that the gap between the mutual information of the resulting scalar channel to $I_{\rm OPT}$ is bounded by $\sim2.51$ dB (or approximately $0.833$ bits per symbol). It is also shown that the bound is tight when both the number of antennas $M$, as well as the $\rm{SNR}$, go to infinity. 

The bound derived in \cite{Narula} is the starting point of the present work. We will employ the scalar coding schemes which were developed since the publication of \cite{Narula} to tighten the bound to $\sim1.1$ dB (or 0.39 bits/symbol). The concept of converting MIMO channel to scalar channel using orhtogonal space-time block codes was presented in \cite{Ganesan}. In order to compare the performance of non-orthogonal methods (whose performance depends on the specific channel), we employ them in conjunction with IR-BF, resulting in an NDO scheme.

Applying IR-BF results in a time-varying SISO channel. Since we assume that the MISO channel is constant over long coded transmission blocks, we take the ergodic mutual information of the resulting SISO channel as the performance metric.  


\section{Review of Known Results}
\label{sec:Review of Known Results}

\subsection{Randomized beamforming}
We begin by reviewing ``standalone" randomized beamforming as presented in \cite{Narula}. In terms of Fig.~\ref{fig:scalarScheme}, this corresponds to having no linear/post processing beyond randomized beamforming.

IR-BF implements a (pseudo) random time-weighting strategy to transform the vector-input channel into a scalar-input channel with
\begin{em} time-varying \end{em} SNR. 
A vector input is generated by multiplying a scalar input by a complex-valued unit-magnitude vector $\boldsymbol{b}_{1,t}$, chosen randomly and uniformly over the surface of the $M$-dimensional complex unit sphere, which is known also to the receiver.
The beamforming vector $\boldsymbol{b}_{1,t}$ can also be interpreted as the first vector of a matrix $\svv{B}_t$, which is drawn from a ``circular unitary ensemble'' (see e.g. \cite{Mehta}) every channel use.
Thus, the transmitted vector at time instance $t$, is
\begin{eqnarray}
\boldsymbol{\tilde{x}}_t= \boldsymbol{b}_{1,t} x_t = \svv{B}_t\begin{bmatrix} 1 & 0 & ... & 0\end{bmatrix}^T x_t
\label{irbfonly}
\end{eqnarray}
where $x_t$, the ``scalar input", is a symbol from a codeword.
The resulting (ergodic) mutual information is
$$I_{\rm RAN}=E_{\boldsymbol{b}_{1,t}}\left[\log_2\left(1+\frac{||\boldsymbol{h}^T\cdot \boldsymbol{b}_{1,t}||^2\eps_s}{N_0}\right)\right].$$
Since $\boldsymbol{b}_{1,t}$ is isotropically distributed, it follows that the mutual information is NDO.
Therefore, combining (i.e., concatenating) any scalar coding scheme with IR-BF, yields a scheme with NDO performance.

We consider the combination of IR-BF with linear modulation schemes. Thus, the modulation  (\ref{irbfonly}), is generalized to be of the form
\begin{align}
\svv{\tilde{X}_k}=\svv{B}_k\svv{P}\svv{X}_k,
\end{align}
where $k$ represents the modulation block index, $\svv{P}$ is a $M\times M$ unitary matrix and $\svv{X}_k$ is a $M\times M$ matrix which corresponds to a linear space-time modulation represents the $k$'th modulation block generated from $M$ inputs. 
Generalizing~(\ref{eq:basicScalarModel}) in this case results in
\begin{eqnarray}
y^{\rm{rec}}_{t} & = & \boldsymbol{h}^T\boldsymbol{x}_{t}+n^{\rm{rec}}_{t}
\label{eq:basicScalarModelForSTBC}
\end{eqnarray}
where $t=kM+l$, $\svv{x}_{t}$ is the $l$'th column of the linear space-time modulation corresponding to the $k$'th modulation block and $y^{\rm{rec}}_{t}$ is the output at time $l$ for the $k$'th modulation block. Equation~(\ref{eq:basicScalarModelForSTBC}) can also be written as
\begin{eqnarray}
{\boldsymbol{y}^{\rm{rec}}_{k}}^T & = & \boldsymbol{h}^T\svv{X}_{k}+{\boldsymbol{n}^{\rm{rec}}_{k}}^T.
\end{eqnarray}

As mentioned above, standalone IR-BF results in a WC gap-to WI mutual information no greater than $\sim2.51$ dB. In the sequel, it is shown that this gap can be reduced by combining (concatenating) IR-BF with further linear pre/post processing as just described. For ease of notation we drop $k$ in the sequel.

\subsection{Alamouti Modulation}
\begin{figure}[p]
      \begin{center}
		 \includegraphics[width=\columnwidth]{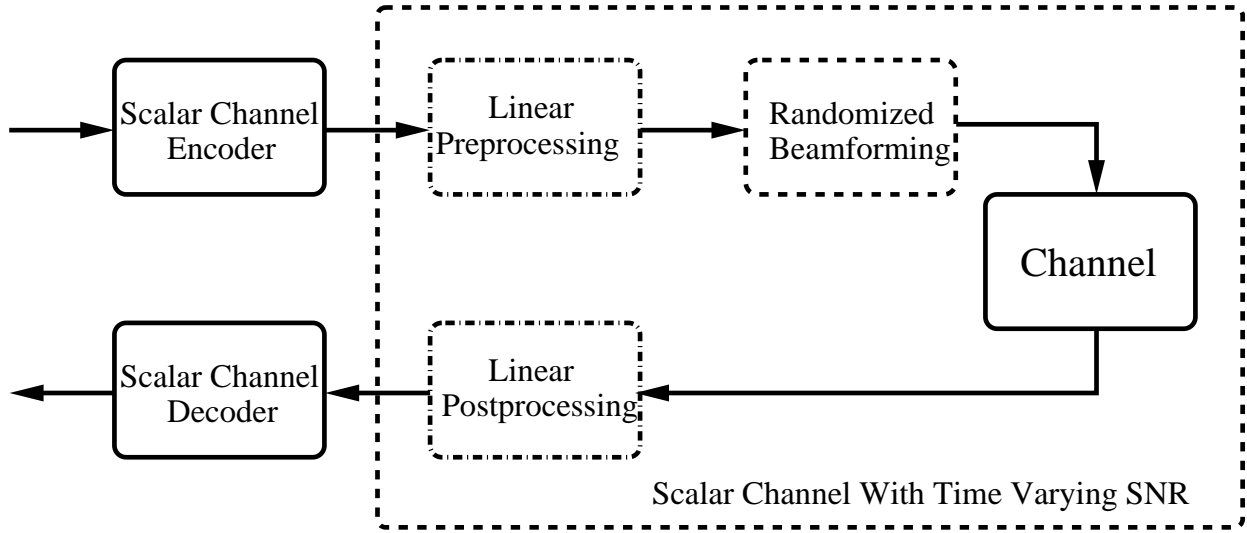}
      \end{center}
\caption{Scalar Coded Antenna System.}
  \label{fig:scalarScheme}
\end{figure}

Alamouti modulation applies to the case of two transmit antennas ($M=2$).
In Alamouti modulation, the linear\footnote{Strictly speaking, Alamouti modulation, as well as QOSTBC, is ``widely linear'', i.e., linear over the reals \cite{Hassibi}.} pre-processing applied to the scalar input, as depicted in Fig.~\ref{fig:scalarScheme}, is
\begin{align}
\svv{X}_{\rm Alamouti}=\frac{1}{\sqrt{2}}\begin{bmatrix} x_1 & x_2^*\\x_2 & -x_1^*\end{bmatrix}. \nonumber
\end{align}
The received symbols can be written as (see e.g. \cite{Badic})
\begin{eqnarray}
\boldsymbol{y}=\svv{H}_{\rm Alamouti}\boldsymbol{x}+\boldsymbol{n},
\end{eqnarray}
where $\boldsymbol{y}=\begin{bmatrix}y^{\rm{rec}}_{1} \\ {y^{\rm{rec}}_{2}}^* \end{bmatrix}$, $\boldsymbol{n}=\begin{bmatrix}n^{\rm{rec}}_{1} \\ {n^{\rm{rec}}_{2}}^* \end{bmatrix}$, $\boldsymbol{x}=\begin{bmatrix}x_1 \\ x_2 \end{bmatrix}$ and $\svv{H}_{\rm Alamouti}$ is the equivalent virtual channel matrix (EVCM)
\begin{align}
\svv{H}_{\rm Alamouti}=\frac{1}{\sqrt{2}}
\begin{bmatrix}
h_1 & h_2 \\
-h_2^* & h_1^*
\end{bmatrix}. \nonumber
\end{align}
Since $\svv{H}_{\rm Alamouti}$ is orthogonal, Alamouti modulation can be optimally demodulated by inverting the EVCM. This operation corresponds to the linear post-processing operation depicted in Fig.~\ref{fig:scalarScheme}. This yields
\begin{align}
\boldsymbol{\bar{y}} = \left(\frac{\sqrt{2}\svv{H}_{\rm Alamouti}^H}{\|\boldsymbol{h}\|}\right) \cdot \boldsymbol{y} = \frac{1}{\sqrt{2}}\begin{bmatrix}\|\boldsymbol{h}\| & 0 \\ 0 & \|\boldsymbol{h}\| \end{bmatrix}\boldsymbol{x}+\boldsymbol{\bar{n}}
\label{ALDEMOD}
\end{align}
where $\bar{\boldsymbol{n}}$ has the same statistics as ${\boldsymbol{n}}$. The resulting scalar channel is
$$y=\frac{{\|\boldsymbol{h}\|}}{\sqrt{2}}x+\bar{n},$$
and its mutual information is
\begin{eqnarray*}
I_{\rm Alamouti}(\SNR,M)=\log_2\left(1+\frac{\|\boldsymbol{h}\|^2}{2} \SNR \right)=I_{\rm OPT}(\SNR).
\end{eqnarray*}
Thus, for the case of two transmit antennas, Alamouti modulation provides an optimal solution to the transmission
problem considered in this work.

\subsection{QOSTBC Modulation}
As mentioned in Section~\ref{sec:scalar coding}, Alamouti modulation is the only member of the OSTBC family that does not sacrifice
the number of utilized degrees of freedom (symbol rate). Several QOSTBC schemes were proposed to circumvent the loss of symbol rate of  OSTBC modulation, at the price of giving up some of the orthogonality.
For simplicity of exposition, we discuss here the case of four transmit antennas. An extension to a larger number of antennas appears in, e.g.,  \cite{Tirkkonen}.

As a representative of QOSTBC schemes, we illustrate the ABBA scheme \cite{Papadias&Foschini}. The linear pre-processing applied to the scalar input, as depicted in Fig.~\ref{fig:scalarScheme}, is thus taken to consist of ABBA modulation
\begin{align}
\svv{X_{{\rm ABBA}}}=\frac{1}{2}
\begin{bmatrix}
x_1 & x_2^* & x_3 & x_4^*\\
x_2 & -x_1^* & x_4 & -x_3^*\\
x_3 & x_4^* & x_1 & x_2^*\\
x_4 & -x_3^* & x_2 & -x_1^*
\end{bmatrix}. \nonumber
\end{align}
As described, e.g., in \cite{Badic}, the received symbols can be written as
 $$
 \boldsymbol{y}=\svv{H}_{\rm ABBA}\boldsymbol{x}+\boldsymbol{n},
 $$
where the EVCM of ABBA is
\begin{align}
\svv{H}_{\rm ABBA}=\frac{1}{2}
\begin{bmatrix}
h_1 & h_2 & h_3 & h_4\\
-h_2^* & h_1^* & -h_4^* & h_3^*\\
h_3 & h_4 & h_1 & h_2\\
-h_4^* & h_3^* & -h_2^* & h_1^*
\end{bmatrix}. \nonumber
\end{align}
Since we confine our attention to linear processing, we consider the linear post-processing (see Fig.~\ref{fig:scalarScheme}) to
consist of a linear MMSE estimator
\begin{align}
\svv{A}_{\rm ABBA}=(\svv{H}_{\rm ABBA}\svv{H}_{\rm ABBA}^H+\SNR^{-1}I)^{-1}\svv{H}_{\rm ABBA}^H.
\label{eq:A_ABBA}
\end{align}
The resulting scalar channel is
\begin{align}
\boldsymbol{\bar{y}} = \svv{A}_{\rm ABBA}\svv{H}_{\rm ABBA}\boldsymbol{x} + \boldsymbol{\bar{n}}
\label{eq:barYabba}
\end{align}
where (with abuse of notation),  $\boldsymbol{\bar{n}}$ is now the resulting filtered noise.
We note that the noise is no longer white.
We assume however that this dependance is not exploited by the scalar code and therefore we measure the performance of the scheme by
\begin{align}
I_{\rm ABBA}(\SNR,\svv{H})=\log_2\left( 1+{\SNR}_{\rm ABBA}(\SNR,\svv{H}_{\rm ABBA}) \right)
\end{align}
where
\begin{align}
{\SNR}_{\rm ABBA}(\SNR,\svv{H}_{\rm ABBA}) = \frac{1}{\{(\svv{H_{\rm ABBA}}^H\svv{H_{\rm ABBA}}\SNR+\svv{I})^{-1}\}_{i,i}}-1.
\end{align}
We note that since the diagonal of $\svv{H_{\rm ABBA}}^H\svv{H_{\rm ABBA}}$ is constant, the above value of ${\SNR}_{\rm ABBA}(\SNR,\svv{H})$ is independent of the value of $i$.
That is, the SNR experienced by all for modulated symbols is the same.
We also note that the $-1$ appears in the formula above since one needs to use the unbiased MMSE estimator when computing SNR (see, e.g., \cite{Cioffi}). The resulting mutual information is
\begin{align}
I_{\rm ABBA}(\SNR,\svv{H})= \log_2\left(\frac{1}{ \left\{(\svv{H_{\rm ABBA}}^H\svv{H_{\rm ABBA}}\SNR+\svv{I})^{-1}\right\}_{1,1}   }   \right).
\end{align}
Concatenating IR-BF with QOSTBC modulation can be interpreted as applying the unitary matrix $\svv{B}$ to the QOSTBC modulation matrix
$$\svv{\tilde{X}}_{{\rm QOSTBC}}=\svv{B}\svv{X}_{{\rm QOSTBC}}.$$
Continue to use ABBA as a representative for QOSTBC modulation, the resulting EVCM of concatenating IR-BF with ABBA can be written as
 \begin{align}
\svv{\tilde H}_{{\rm ABBA}}=\frac{1}{2}
\begin{bmatrix}
\tilde{h}_1 & \tilde{h}_2 & \tilde{h}_3 & \tilde{h}_4\\
-\tilde{h}_2^* & \tilde{h}_1^* & -\tilde{h}_4^* & \tilde{h}_3^*\\
\tilde{h}_3 & \tilde{h}_4 & \tilde{h}_1 & \tilde{h}_2\\
-\tilde{h}_4^* & \tilde{h}_3^* & -\tilde{h}_2^* & \tilde{h}_1^*
\end{bmatrix}, \nonumber
\end{align}
where
\begin{align}
\tilde{h}_1=\boldsymbol{h}^T \boldsymbol{b}_{1} \nonumber \\
\tilde{h}_2=\boldsymbol{h}^T \boldsymbol{b}_{2} \nonumber \\
\tilde{h}_3=\boldsymbol{h}^T \boldsymbol{b}_{3} \nonumber \\
\tilde{h}_4=\boldsymbol{h}^T \boldsymbol{b}_{4}.
\label{eq:tildeh}
\end{align}
The demodulator $\svv{A}_{\rm ABBA}$ is the same as defined in (\ref{eq:A_ABBA}), but now using $\svv{\tilde{H}}_{{\rm ABBA}}$ (per block) in place of $\svv{{H}}_{{\rm ABBA}}$.
We note that the matrix $\svv{B}$ is drawn per QOSTBC block.

The resulting (ergodic) mutual information can be expressed as
\begin{align}
I_{\rm IR-ABBA}(\SNR,M)= E_{\svv{B}}\left[\log_2\left(1+{\SNR}_{\rm ABBA}(\SNR,\svv{\tilde{H}}_{{\rm ABBA}})\right)\right]
\end{align}

We note that since all QOSTBC variants can be translated to each other by applying a linear {\em unitary} transformation \cite{Badic}, it is not difficult to show that, the combination of any of the QOSTBC variants discussed above with IR-BF, yields the same results.

\subsection{TROMBI/STTD-PHOP}
An alternative modulation approach (to QOSTBC) based on Alamouti modulation was proposed in \cite{TROMBI}, where it was named TROMBI. In TROMBI modulation, two identical copies of Alamouti modulation are summed together with pseudo-random phases. The pre-processing of TROMBI can be viewed as applying the following beamforming matrix
$$\svv{X_{\rm TROMBI}}=\svv{P}_{\rm TROMBI} \svv{X}_{\rm Alamouti}$$
where
\begin{align}
\svv{P}_{\rm TROMBI}=
\frac{1}{\sqrt{2}}
\begin{bmatrix}
1 & 0 \\
e^{j\theta_1} & 0\\
0 & 1\\
0 & e^{j\theta_2}
\end{bmatrix}, \nonumber
\end{align}
which results in
\begin{align}
\svv{X}_{\rm TROMBI}=\frac{1}{{2}}
\begin{bmatrix}
x_1 & x_2^* \\
e^{j\theta_1}x_1 & e^{j\theta_1}x_2^* \\
x_2 & -x_1^* \\
e^{j\theta_2}x_2 & -e^{j\theta_2}x_1^*
\end{bmatrix}.
\label{eq:TROMBI_NO_IR}
\end{align}
The EVCM of TROMBI is
\begin{align}
\svv{H}_{\rm TROMBI}=\frac{1}{\sqrt{2}}
\begin{bmatrix}
{h}_{1,{\rm TROMBI}} & {h}_{2,{\rm TROMBI}} \\
-{h}_{2,{\rm TROMBI}}^* & {h}_{1,{\rm TROMBI}}^* \\
\end{bmatrix}
\end{align}
where
\begin{align}
{h}_{1,{\rm TROMBI}}=\frac{1}{\sqrt{2}}\left(h_1+h_2\cdot e^{j\theta_1}\right) \nonumber \\
{h}_{2,{\rm TROMBI}}=\frac{1}{\sqrt{2}}\left(h_3+h_4\cdot e^{j\theta_2}\right)
\label{eq:TrombiWithouhtRandChannels}
\end{align}
and these can be used in the Alamouti demodulator in place of the actual channel coefficients. The (ergodic) mutual information of TROMBI can be written as
$$I_{\rm TROMBI}(\SNR,\boldsymbol{h}) = E_{\theta_1,\theta_2}\left[\log_2\left(1+\SNR\frac{\|\boldsymbol{h}_{\rm TROMBI}\|^2}{2} \right)\right]$$
where $\boldsymbol{h}_{\rm TROMBI}$ is given in (\ref{eq:TrombiWithouhtRandChannels}).

In the notations of Fig.~\ref{fig:scalarScheme}, the linear pre-processing is a cascade of Alamouti modulation and the beamforming operation $\svv{P}_{\rm TROMBI}$. The linear post-processing operation amounts to applying the Alamouti demodulator (\ref{ALDEMOD}) to the equivalent channel, i.e., multiplying the received symbols by ${\sqrt{2}\svv{H}_{\rm TROMBI}^H}/{\|\boldsymbol{h_{\rm TROMBI}}\|^2}$.

We note that the performance of TROMBI depends on the specific realization of the channel and hence it is not NDO.\footnote{For example, in case of $h_1=h_2=0,h_3=h_4=\sqrt{c}$ , since the transmitted power is equally divided between the two ``virtual'' antennas, the performance will be lower than for the case where $h_1=h_2=h_3=h_4=\sqrt{\frac{c}{2}}$ (whereas in both cases $\|\boldsymbol{h}\|^2$ is the same).} We further note that TROMBI modulation applies random phases rather than general orthogonal directions as done in IR-BF.

Combining TROMBI with IR-BF yields the following transmitted matrix
$$\svv{\tilde{X}}_{{\rm{IR-TROMBI}}}=\svv{B} P_{\rm{TROMBI}} \svv{X}_{{\rm{Alamouti}}}.$$
The EVCM can be written as
\begin{align}
\svv{H}_{{\rm TROMBI}}=
\frac{1}{\sqrt{2}}
\begin{bmatrix}
\tilde{h}_{1,{\rm TROMBI}} & \tilde{h}_{2,{\rm TROMBI}} \\
-\tilde{h}_{2,{\rm TROMBI}}^* & \tilde{h}_{1,{\rm TROMBI}}^* \\
\end{bmatrix}
\end{align}
where
\begin{align}
\tilde{h}_{1,{\rm TROMBI}}=\frac{1}{\sqrt{2}}\left(\tilde{h}_1+\tilde{h}_2\cdot e^{j\theta_1}\right) \nonumber \\
\tilde{h}_{2,{\rm TROMBI}}=\frac{1}{\sqrt{2}}\left(\tilde{h}_3+\tilde{h}_4\cdot e^{j\theta_2}\right) \nonumber
\end{align}
and $\boldsymbol{\tilde{h}}$ is defined in (\ref{eq:tildeh}).
We note that the matrix $\svv{B}$ should be drawn per Alamouti block.

The resulting (ergodic) mutual information is
\begin{align}
I_{\rm IR-TROMBI}(\SNR,M) = E_{\theta_1,\theta_2,\svv{B}}\left[\log_2\left(1+\SNR\frac{\|\boldsymbol{\tilde{h}}_{\rm TROMBI}\|^2}{2} \right)\right]. \nonumber
\end{align}


\subsection{Comparison Of Different Methods}
\begin{figure}
      \begin{center}
		 \includegraphics[width=\columnwidth]{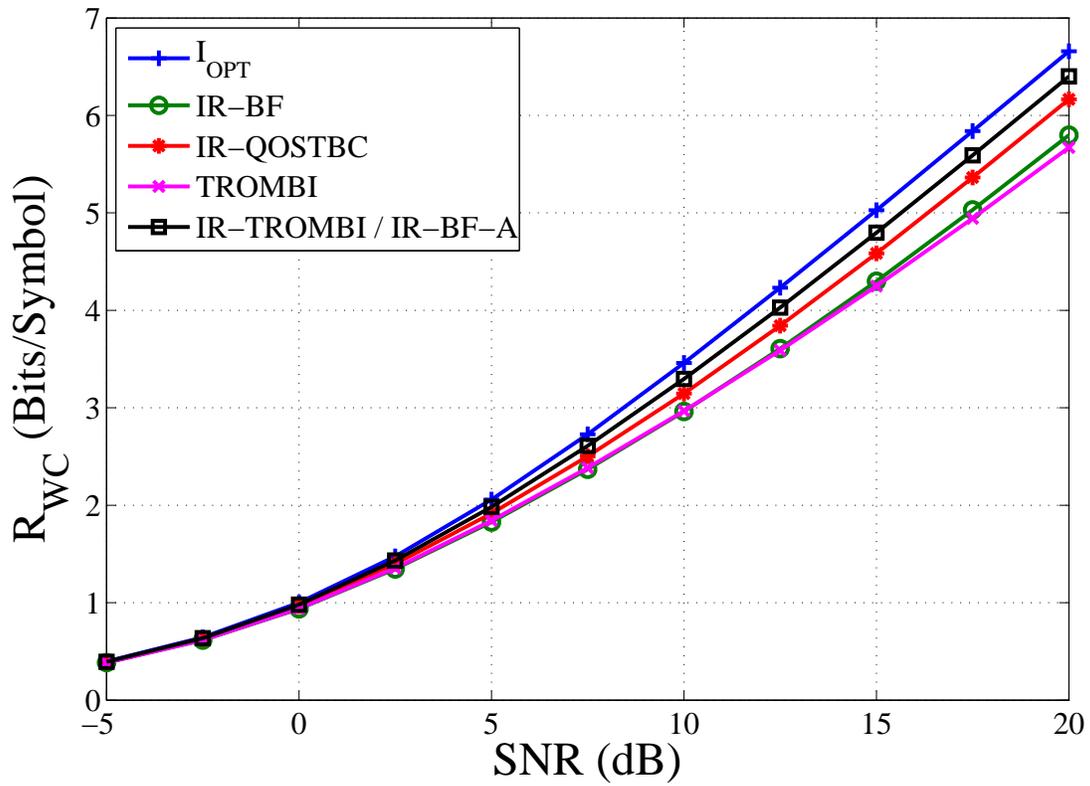}
      \end{center}
\caption{Guaranteed performance of quasi-orthogonal modulation for four antennas.}
  \label{fig:CombiningQuasiOrthoWithRandBeamForm}
\end{figure}

The ergodic mutual information obtained by combining IR-BF with QOSTBC/TROMBI is plotted in Fig.~\ref{fig:CombiningQuasiOrthoWithRandBeamForm}. As a representative for QOSTBC, IR-ABBA was simulated. For reference, the performance of standalone IR-BF is shown, as well as the WC performance of TROMBI without IR-BF. We see that combining QOSTBC with IR-BF improves the mutual information with respect to IR-BF only, and combining TROMBI with IR-BF results in the best WC performance among the considered schemes.




\section{Randomly Beamformed Alamouti}
\label{sec:RBF}
Thus far, the modulation method that results in the best WC performance is that of IR-TROMBI. We note that in this construction, we actually apply two layers of randomization to the data. When concatenated with IR-BF, the (partial) randomization of TROMBI becomes redundant, i.e., setting $\theta_1,\theta_2$ to any constant value yields the same performance. We further note that choosing any orthogonal two-column projection matrix will yield the same performance.

\begin{figure}
\begin{center}
\begin{psfrags}
\psfrag{a1}[][][1]{$\overbrace{x_{2},x_{1}}$}
\psfrag{a3}[][][1]{$\overbrace{x_{4},x_{3}}$}
\psfrag{b1}[][][1]{$x_{1}$}
\psfrag{b2}[][][1]{$x_{2}$}
\psfrag{c1}[][][1]{$-x_{2}^*$}
\psfrag{c2}[][][1]{$x_{1}^*$}
\psfrag{e1}[][][1]{$\boldsymbol{b}_{1}$}
\psfrag{e2}[][][1]{$\boldsymbol{b}_{2}$}
\psfrag{d1}[][][1]{$\boldsymbol{\frac{1}{\sqrt{2}}}$}
\psfrag{d2}[][][1]{$\boldsymbol{\frac{1}{\sqrt{2}}}$}
\psfrag{f1}[][][1]{$\boldsymbol{\tilde{X}}$}
\includegraphics[width=\columnwidth]{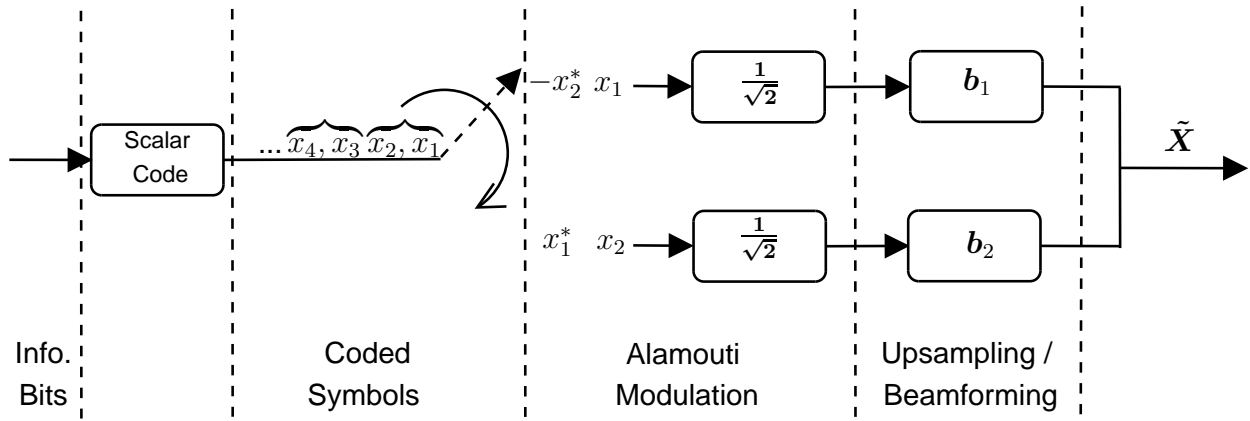}
\end{psfrags}
\end{center}
\caption{Randomly beamformed Alamouti modulation.}
\label{fig:NewSchemeHighLevel}
\end{figure}

We now show that one can combine these two layers of randomization into a single layer that performs projection onto two random orthogonal ``directions''. We refer to this method as isotropic randomly beamformed Alamouti (IR-BF-A). IR-BF-A is illustrated in Fig.~\ref{fig:NewSchemeHighLevel}. We designate by $\boldsymbol{b}_{1}$ and $\boldsymbol{b}_{2}$ the first two columns of a random unitary matrix $\svv{B}$ drawn from a circular unitary ensemble (see $\cite{Narula}$). The transmitted signal is
\begin{align}
\svv{\tilde{X}}_{{\rm{IR-BF-A}}}=\svv{B}
\begin{bmatrix} 1 & 0 & 0 & ... & 0 \\
0 & 1 & 0 & ... & 0 \end{bmatrix}^T \svv{X}_{{\rm Alamouti}}. \nonumber
\end{align}
This can be viewed as if the equivalent channel that the data streams experience is
\begin{align}
\begin{bmatrix} \tilde{h}_1 & \tilde{h}_2 \end{bmatrix} =& \begin{bmatrix} h_1 & h_2 & ... & h_M \end{bmatrix} \svv{B}
\begin{bmatrix} 1 & 0 & 0 & ... & 0 \\
0 & 1 & 0 & ... & 0 \end{bmatrix}^T
\label{eq:hTildeOfBFA}
\end{align}


We have now reduced the $1\times M$ system to an effective $1\times2$ system with channel gains $\tilde{h}_1,\tilde{h}_2$, onto which we may next apply Alamouti modulation. The (ergodic) mutual information of the scheme is
\begin{align}
I_{\rm{IR-BF-A}}(\SNR,M)=E_{\svv{B}}\left[\log_2\left(1+\SNR\frac{\|\boldsymbol{\tilde{h}}\|^2}{2}\right)\right].
\label{eq:IR-BF-A}
\end{align}
We note that because of the random rotation matrix $\svv{{B}}$, the mutual information is NDO.

Fig.~\ref{fig:CombiningQuasiOrthoWithRandBeamForm} depicts the corresponding mutual information as a function of $\SNR$. As explained above, the curve of IR-TROMBI coincides with the IR-BF-A curve. It exceeds IR-ABBA and unlike QOSTBC with a linear MMSE front end,\footnote{In fact, simulations show that performance of IR-BF-A is nearly identical to that of ABBA with optimal (non-linear) front end.} the front end for IR-BF-A {does not require matrix inversion}.

To the best of the authors knowledge, IR-BF-A yields the best WC performance of scalar coding known to date. Thus, by upper bounding the gap-to WI mutual information of IR-BF-A, we obtain the tightest known bound on the gap-to WI mutual information of scalar coding.

\begin{figure}
      \begin{center}
		 \includegraphics[width=\columnwidth]{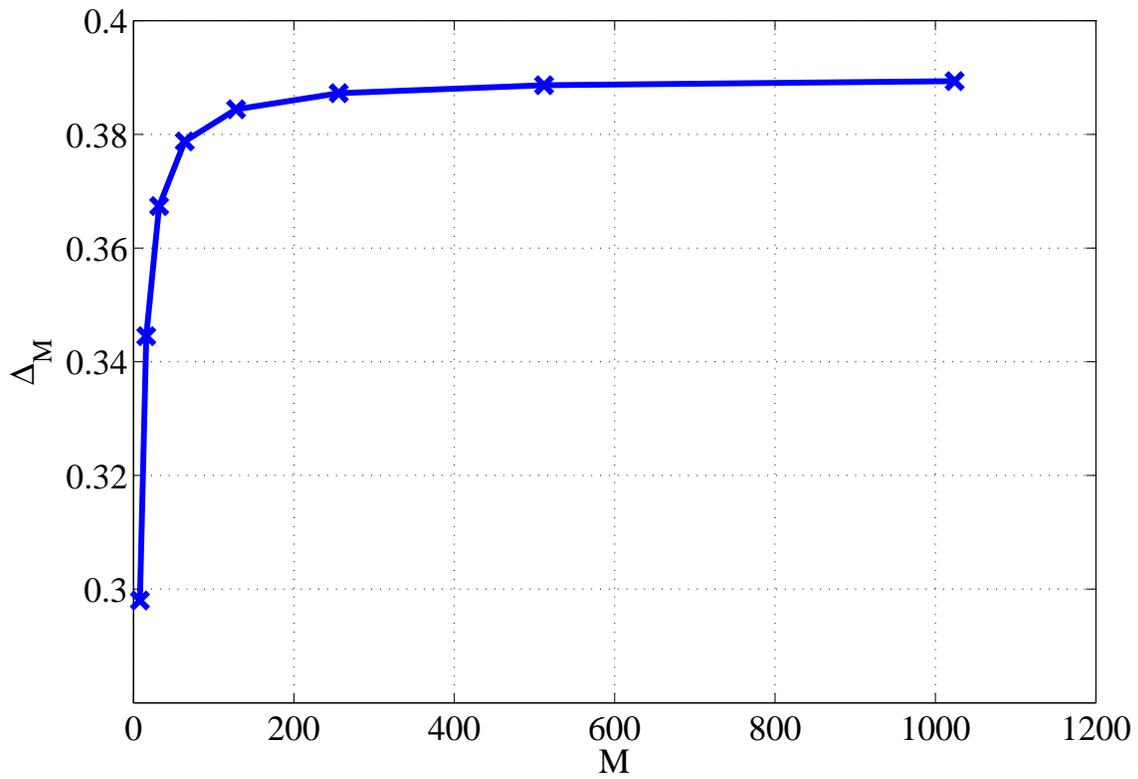}
      \end{center}
\caption{IR-BF-A gap to WI mutual information as a function of $M$.}
  \label{fig:deltaAsAFunctionOfMInfSNR}
\end{figure}

To that end, denote the gap between IR-BF-A to the WI mutual information by
$$\Delta(\SNR,M)=I_{\rm OPT}(\SNR,M)-I_{\rm{IR-BF-A}}(\SNR,M).$$
The following two lemmas are proved in the Appendix.
\vspace{2mm}
\begin{lemma}
The gap-to WI mutual information $\Delta(\SNR,M)$ is given by
\begin{align}
\Delta(\SNR,M)=&(M-1)(M-2)\displaystyle\int_0^{1}\log_2\left(\frac{1+{\SNR}}{1+M\frac{\SNR}{2}u^2}\right) \nonumber \\
& \times u^3(1-u^2)^{M-3} du.
\label{eq:FullDelta}
\end{align}
\end{lemma}
\vspace{2mm}
\begin{lemma}
$\Delta(\SNR,M)$ is monotonically increasing with $\SNR$.
\end{lemma}
\vspace{2mm}
Denote $\Delta_M=\Delta(\infty,M)$. Using (\ref{eq:FullDelta}), we obtain the asymptotic gap
\begin{align}
\Delta_M&=\lim_{\SNR \to \infty}\Delta(\SNR,M) \nonumber \\
&=(M-1)(M-2)\displaystyle\int_0^{1}\log_2\left(\frac{2}{u^2\ M}\right)u^3(1-u^2)^{M-3}du.
\label{eq:deltaM}
\end{align}
This asymptotic gap $\Delta_M$ as a function of the number of transmit antennas is illustrated in Fig.~\ref{fig:deltaAsAFunctionOfMInfSNR}.
As can be observed numerically, the maximal gap is attained when $M\to\infty$. We denote the maximal gap by
\begin{align}
\Delta=\lim_{M\to\infty}\Delta_M.
\label{eq:deltaForTheorem}
\end{align}
Combining the above we can show that substituting $z=Mu^2$ in (\ref{eq:deltaM}), we obtain
\begin{align}
\Delta_M=\displaystyle\int_0^{M}\log_2\left(\frac{2}{z} \right)\frac{(M-1)(M-2)}{M^2}z\left(1-\frac{z}{M}\right)^{M-3}dz. \nonumber
\end{align}
Taking ${M\to\infty}$ yields
\begin{align}
\Delta = \displaystyle\int_0^{\infty}\log_2\left(\frac{2}{z}\right)ze^{-z}dz, \nonumber
\end{align}
which may also be written as
\begin{align}
 \Delta & = \displaystyle\int_0^{\infty}ze^{-z}dz + \displaystyle\int_0^{\infty}\log_2\left(\frac{1}{z}\right)ze^{-z}dz \nonumber \\
& = 1 + \displaystyle\int_0^{\infty}\log_2\left(\frac{1}{z}\right)ze^{-z}dz. \nonumber
\end{align}
Using integration by parts we get
\begin{align}
 \Delta & = 1- \log_2\left(\frac{1}{z}\right)ze^{-z}|_0^{\infty}-\displaystyle\int_0^{\infty}\left(\log_2(e)+\log_2(z)\right)e^{-z}dz \nonumber \\
& = 1 - (\log_2(e) + \displaystyle\int_0^{\infty}\log_2(z)e^{-z}dz)   \nonumber \\
& = 1- (\log_2(e) - \gamma) \approx0.39. \nonumber
\end{align}
where $\gamma$ is Euler's constant which is defined as $$\gamma=\displaystyle\int_0^{\infty}\log_2\left(\frac{1}{z}\right)e^{-z}dz
\approx0.833.$$

We conclude that the maximal gap-to WI mutual information as defined in (\ref{eq:deltaForTheorem}) satisfies
\begin{align}
\Delta = 1- \left( \log_2 \left( e \right) -\gamma \right)\approx0.39~{\rm bits/symbol}. \nonumber
\end{align}

\begin{figure}
      \begin{center}
		 \includegraphics[width=\columnwidth]{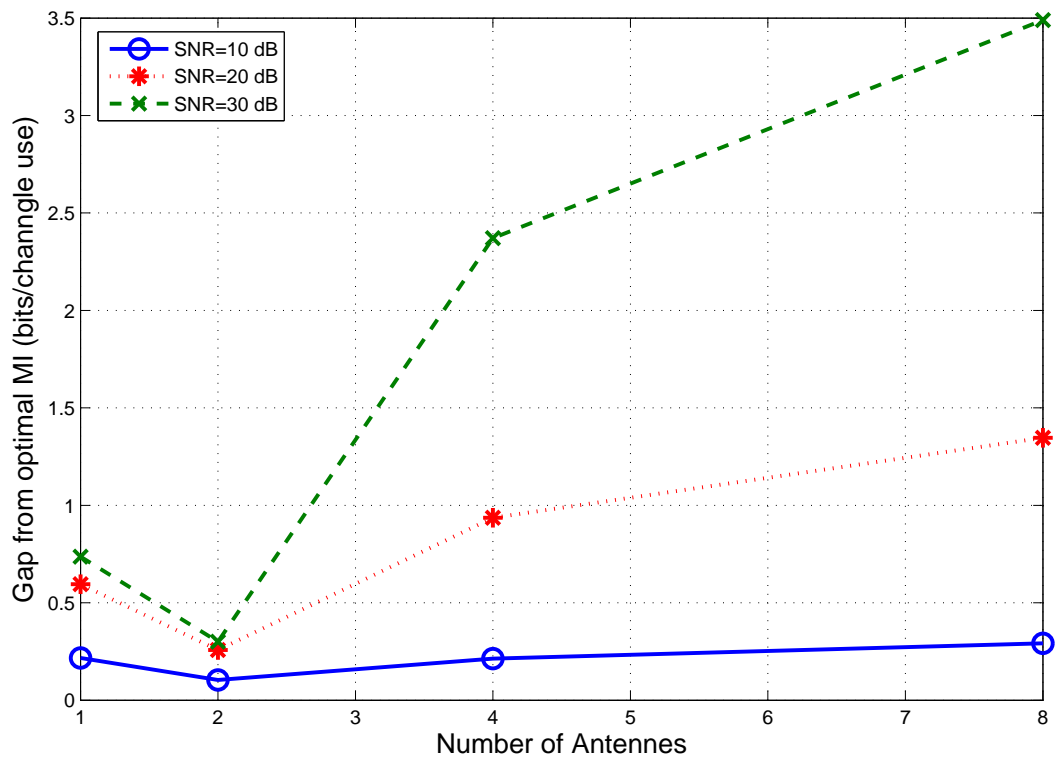}
      \end{center}
\caption{Gap to WI mutual information for various ``virtual'' number of antennas.}
  \label{fig:misoToNiso}
\end{figure}

A question the arises following our analysis is whether using more virtual antennas (in conjunction with OSTBC modulation) may result in a smaller gap to the WI mutual information. As discussed above, any OSTBC modulation beyond the case of Alamouti suffers from an inherent loss in symbol rate. This suggests that using two ``virtual antennas'' will be the best at least at high SNR. Numerical simulation shows that this is indeed the case at all SNR values. In Fig.~\ref{fig:misoToNiso} below we simulated eight transmit antennas, and we present the gap from the WI mutual information for various number of ``virtual'' antennas.

We compare reduction to single ``virtual'' antenna (\cite{Narula}), to two ``virtual'' antennas (IR-BF-A) and to four and eight ``virtual'' antennas. The reduction to four and eight was done by extending the method developed in the paper, i.e., projecting the original channel onto more orthogonal vectors from the matrix $\svv{B}_k$ (which is drawn from a circular unitary ensemble). The instantaneous mutual information of the resulting ``virtual'' MISO channel is the mutual information of optimal OSTBC for this number of antennas (\cite{Tarokh}). The gap is calculated from the ergodic mutual information attained by this method of projection and modulation. Using two ``virtual'' antennas results with the smallest gap.

\section{Conclusion}
\label{sec:conc}

In this paper, we compared scalar coding schemes and analyzed the worst-case performance of these schemes. Combination of Alamouti and isotropic randomized beamforming was identified as the scalar coded scheme achieving the best performance known to date and the gap-to WI mutual information of scalar-coded channels was shown to be no more than $0.39$ bits/symbol.

\section{Appendix}
\setcounter{lemma}{0}
\subsection{Proof of Lemma 1}
The explicit formula for the gap-to WI mutual information is
\begin{align}
\Delta(\SNR,M)=I_{\rm{OPT}}(\SNR)-I_{\rm{IR-BF-A}}(\SNR,M).
\label{eq:gapDefintion}
\end{align}
The WI mutual information is given in (\ref{eq:OPTwithChanNorm1}).
Substituting~(\ref{eq:hTildeOfBFA}) into~(\ref{eq:IR-BF-A}), we get an explicit expression for $I_{\rm{IR-BF-A}}$
\begin{align}
& I_{\rm{IR-BF-A}}(\SNR,M) = \nonumber \\
& E _{\svv{B}}\left[\log_2\left(1+\frac{\SNR}{2}\left(|\boldsymbol{h}^T \boldsymbol{b}_{1}|^2+|\boldsymbol{h}^T \boldsymbol{b}_{2}|^2\right)\right)\right].
\label{eq:EQ_BFA_basic}
\end{align}
Now, define $\bar{h_i}=\frac{h_i}{\sqrt{M}}$. Since $\frac{\|\boldsymbol{h}\|^2}{M}=1$ it follows that $\|\boldsymbol{\bar{h}}\|^2=1$. Substituting $h_i=\bar{h_i}\sqrt{M}$ in~(\ref{eq:EQ_BFA_basic}), we get
\begin{align}
& I_{\rm{IR-BF-A}}(\SNR,M) = \nonumber \\
& E_{\svv{B}}\left[\log_2\left(1+M\frac{\SNR}{2}\left(|\boldsymbol{\bar{h}}^T \boldsymbol{b}_{1}|^2+|\boldsymbol{\bar{h}}^T \boldsymbol{b}_{2}|^2\right)\right)\right].
\label{eq:EQ_BFA1}
\end{align}
Denoting
\begin{align}
r=\sqrt{|\boldsymbol{\bar{h}}^T \boldsymbol{b}_{1}|^2+|\boldsymbol{\bar{h}}^T \boldsymbol{b}_{2}|^2},
\label{eq:radiusDef}
\end{align}
equation (\ref{eq:EQ_BFA1}) becomes
\begin{align}
I_{\rm{IR-BF-A}}(\SNR,M) = E_{\svv{B}}\left[\log_2\left(1+M\frac{\SNR}{2}r^2\right)\right].
\label{eq:IR-BF-A_withR}
\end{align}
We characterize the pdf of $r$ using the following proposition which appears in \cite{UnitMat}.
\begin{proposition}
\label{prop:prop1}
The pdf of a projection of a vector which is uniformly distributed over the surface of a (real) $M$-dimensional sphere of unit radius, onto a (real) $N$-dimensional linear subspace is given by
\begin{align}
P_{M,N}(u)=\frac{2\Gamma(\frac{M}{2})}{\Gamma(\frac{N}{2})\Gamma(\frac{M-N}{2})}u^{N-1}(1-u^2)^{(M-N-2)/2}.
\label{eq:orthPtoj}
\end{align}
\end{proposition}
\vspace{2mm}
In IR-BF-A modulation, since $\svv{B}$ is a random unitary matrix drawn from a circular unitary ensemble, it follows that $\boldsymbol{b}_{1}$ and $\boldsymbol{b}_{2}$ are two orthogonal vectors, distributed uniformly over the surface of the (complex) $M$-dimensional unit sphere. From symmetry, this is equivalent to fixing $\boldsymbol{b}_{1}=\begin{bmatrix} 1 & 0 & ... & 0\end{bmatrix}$ and $\boldsymbol{b}_{2}=\begin{bmatrix} 0 & 1 & ... & 0 \end{bmatrix}$ and taking $\boldsymbol{h}$ to be uniformly distributed over the surface of a (complex) M-dimensional sphere with radius $||\boldsymbol{\bar{h}}||=1$.

This means that $f_r(u)$ is the pdf of a projection of a vector which is uniformly distributed over the surface of a (complex) M-dimensional sphere of unit radius, onto a (complex) two-dimensional plane.
Using Proposition~\ref{prop:prop1}, we get
\begin{align}
f_r(u)=P_{2M,4}(u)=2(M-1)(M-2)u^3(1-u^2)^{M-3},
\label{eq:pdfOfr}
\end{align}
where we substitute $2M$ and $4$ in (\ref{eq:orthPtoj}) since we're projecting from $M$ complex dimensions to two complex dimensions.
The explicit expression for $I_{\rm{IR-BF-A}}$ is
\begin{align}
& I_{\rm{IR-BF-A}}(\SNR,M)= E_{\svv{B}}\left(\log_2\left(1+M\frac{\SNR}{2}r^2\right)\right) = \nonumber \\
& 2(M-1)(M-2)\displaystyle\int_0^{1} \log_2\left(1+M\frac{\SNR}{2}u^2 \right) \nonumber \\
& \times u^3(1-u^2)^{M-3} du.
\label{eq:EQ_BFA}
\end{align}
Substituting (\ref{eq:EQ_BFA}) and (\ref{eq:OPTwithChanNorm1}) in (\ref{eq:gapDefintion}) gives the explicit formula for the gap
\begin{align}
\Delta(\SNR,M) = & 2(M-1)(M-2)\displaystyle\int_0^{1}\log_2\left(\frac{1+{\SNR}}{1+M\frac{\SNR}{2}u^2}\right)\nonumber \\
& \times u^3(1-u^2)^{M-3}du.\nonumber
\end{align}
\subsection{Proof of Lemma 2}

For the sake of this lemma, we don't need the explicit expressions developed in Lemma 1. Rather, we consider directly the definition of the gap-to WI mutual information. Combining (\ref{eq:OPTwithChanNorm1}) and (\ref{eq:IR-BF-A_withR}), we observe that the gap-to WI mutual information of IR-BF-A is given by
\begin{align}
& \Delta(\SNR,M)=I_{\rm{OPT}}(\SNR)-I_{\rm{IR-BF-A}}(\SNR,M) = \nonumber \\
& \log_2\left(1+{\SNR}\right)-E_{\svv{B}}\left[\log_2\left(1+M \frac{\SNR}{2}r^2\right)\right]. \nonumber
\end{align}
where $r$ is defined in (\ref{eq:radiusDef}).
Integrating (\ref{eq:pdfOfr}) we get $E_{\svv{B}}\left[{r^2}\right]=\frac{2}{M}$. Differentiating $\Delta(\SNR,M)$ and exchanging the order of differentiation and expectation yields
\begin{align}
& \frac{d\Delta(\SNR,M)}{d\SNR} = \frac{1}{1+\SNR}-E_{\svv{B}}\left[\frac{\frac{M r^2}{2}}{1+\frac{\SNR M}{2}r^2} \right] \nonumber \\
& = \frac{1}{1+\SNR}-\frac{1}{\SNR}\left(1-E_{\svv{B}}\left[\frac{1}{1+\frac{\SNR M}{2}r^2} \right] \right) \nonumber.
\end{align}
Applying Jensen's inequality, we get
$$
E_{\svv{B}}\left[\frac{1}{1+\frac{\SNR M}{2}r^2} \right]\geq \frac{1}{E_{\svv{B}}\left[1+\frac{\SNR M}{2}r^2\right]},
$$
which in term implies that
\begin{align}
\frac{d\Delta(\SNR,M)}{d\SNR} &\geq \frac{1}{1+\SNR}-\frac{1}{\SNR}\left(1-\frac{1}{E_{\svv{B}}\left[1+\frac{\SNR M}{2}r^2\right]}\right) \nonumber \\
&= \frac{{1}}{1+{\SNR}}-\frac{1}{\SNR}\left(1-\frac{1}{1+\SNR}\right)=0. \nonumber
\end{align}
It follows that the gap is increasing with the SNR.

\end{document}

%% file: rv_defs.tex
\DeclareMathAlphabet{\mathbsf}{OT1}{cmss}{bx}{n}
\DeclareMathAlphabet{\mathssf}{OT1}{cmss}{m}{sl}
\DeclareMathAlphabet{\mathcsf}{OT1}{cmss}{sbc}{n}

\newcommand{\svv}[1]{\mathbf{#1}}

\DeclareSymbolFont{bsfletters}{OT1}{cmss}{bx}{n}  
\DeclareSymbolFont{ssfletters}{OT1}{cmss}{m}{n}
\DeclareMathSymbol{\bsfGamma}{0}{bsfletters}{'000}
\DeclareMathSymbol{\ssfGamma}{0}{ssfletters}{'000}
\DeclareMathSymbol{\bsfDelta}{0}{bsfletters}{'001}
\DeclareMathSymbol{\ssfDelta}{0}{ssfletters}{'001}
\DeclareMathSymbol{\bsfTheta}{0}{bsfletters}{'002}
\DeclareMathSymbol{\ssfTheta}{0}{ssfletters}{'002}
\DeclareMathSymbol{\bsfLambda}{0}{bsfletters}{'003}
\DeclareMathSymbol{\ssfLambda}{0}{ssfletters}{'003}
\DeclareMathSymbol{\bsfXi}{0}{bsfletters}{'004}
\DeclareMathSymbol{\ssfXi}{0}{ssfletters}{'004}
\DeclareMathSymbol{\bsfPi}{0}{bsfletters}{'005}
\DeclareMathSymbol{\ssfPi}{0}{ssfletters}{'005}
\DeclareMathSymbol{\bsfSigma}{0}{bsfletters}{'006}
\DeclareMathSymbol{\ssfSigma}{0}{ssfletters}{'006}
\DeclareMathSymbol{\bsfUpsilon}{0}{bsfletters}{'007}
\DeclareMathSymbol{\ssfUpsilon}{0}{ssfletters}{'007}
\DeclareMathSymbol{\bsfPhi}{0}{bsfletters}{'010}
\DeclareMathSymbol{\ssfPhi}{0}{ssfletters}{'010}
\DeclareMathSymbol{\bsfPsi}{0}{bsfletters}{'011}
\DeclareMathSymbol{\ssfPsi}{0}{ssfletters}{'011}
\DeclareMathSymbol{\bsfOmega}{0}{bsfletters}{'012}
\DeclareMathSymbol{\ssfOmega}{0}{ssfletters}{'012}